# Training and Innovation in Italian Manufacturing Firms

Davide Antonioli[*†], Elisa Chioatto[*†], Giovanni Guidetti[‡1], Riccardo Leoncini[‡§], Mariele Macaluso[‖]

*Department of Economics and Management, University of Ferrara, Italy*
*† SEEDS – Sustainability, Environmental Economics and Dynamic Studies, Italy*
*‡ Department of Legal Studies, University of Bologna, Italy*
*§ IRCrES-CNR, Milan, Italy*
*‖ ESOMAS Department, University of Turin, Italy*

## Abstract

This paper analyses how firms' skill development strategies affect their propensity to introduce innovation. We develop an adjustment-cost framework that links human capital theory and institutionalist and evolutionary approaches, considering innovation as an activity that entails costs in labour adjustment arising either from the training activities of workers or the recruitment of skilled employees. Using a two-wave panel of Italian manufacturing firms observed in 2017-2018 and 2019-2020, we analyse firms' adoption of total, product, process, and circular innovation as a function of internal training practices and of external skills acquisition. Overall, the empirical analysis confirms the expected positive relationship between training and innovation, while also revealing important nuances in the workforce upskilling strategies required for different types of innovation. Moreover, while training activities and skills development are essential across all forms of innovation, our findings indicate that internal training is particularly effective in supporting the implementation of circular innovations. By contrast, external recruitment appears to be consistently necessary whenever innovations are introduced, regardless of their type.

## 1. Introduction

Skills play a central role in firms' performance and in their ability to adapt to technological change.

The relationship between skills and innovation is inherently bidirectional: innovation calls for new skills, but skill endowments also condition firms' ability to adopt and exploit new technologies. On the one hand, technological and organizational innovations typically require firms to perform new tasks or to recombine existing ones, thereby increasing the demand for specific skill sets. On the other hand, firms' skill endowment and skill-development strategies determine their capacity to introduce different forms of innovation. In both cases, the link between skills and innovation operates through adjustments in firms' workforce composition. Firms can train their existing workforce, hire new workers with appropriate skills, or combine both strategies.

---

[1] Corresponding author: Giovanni Guidetti, g.guidetti@unibo.it

While these adjustment mechanisms are often implicitly assumed in the literature, they involve non-negligible costs and organizational constraints that may affect firms' innovation decisions. Little attention has been paid to how skill endowments and workforce-development strategies, mediated by these adjustment costs, influence firms' innovation outcomes. This paper directly attempts to addresses this gap.

Most economic studies approach skill formation through human capital theory, focusing on training decisions and productivity outcomes. Although this framework is well suited to analyse labour market returns to skills, it provides limited insights into how skills affect firms' innovative behaviour. In contrast, institutionalist and evolutionary approaches emphasize the role of organizational learning, absorptive capacity, and dynamic capabilities in shaping innovation, but often lack a clear representation of the cost mechanisms through which skills enter firms' innovation decisions.

The aim of this paper is to bridge these perspectives by introducing an adjustment cost framework that links firms' skill endowments to their propensity to innovate. Innovation is modelled as an activity that entails expected labour adjustment costs, arising from the need to train incumbent workers and/or to recruit new employees with appropriate skills. Following Hamermesh (1995), these adjustment costs are assumed to be lumpy and to arise whenever firms modify their employment structure to implement technological or organizational change. From an ex-ante perspective, firms with richer skill endowments, either through trained employees or through access to highly skilled labour, face lower expected adjustment costs and are therefore more likely to introduce innovation. This framework provides an interpretative bridge between the human capital approach, which emphasizes formal and measurable training, and the evolutionary perspective, which highlights informal learning, skill complementarities, and tacit knowledge.

The empirical analysis focuses on Italy, a country characterized by a production structure dominated by small and medium-sized enterprises (SMEs), which rely primarily on incremental and embodied forms of innovation. Using a unique two-wave dataset of Italian SMEs drawn from national surveys conducted in 2017-2018 and 2019-2020, we model firms' innovation outcomes, distinguishing between total, process, product, and circular innovations, as a function of their skill endowments and skill-development strategies, proxied by the presence of trained employees and by the employment of high-skilled workers. More specifically, the paper aims at examining how firms' skill-development strategies, either internal training or the recruitment of skilled employees, affect their propensity to introduce different forms of innovation, and whether different types of innovation rely on distinct combinations of these strategies.

The paper is organised as follows. *Section 2* provides a literature review to identify the main research questions, which are subsequently discussed in *Section 3*. *Section 4* presents the data and the empirical model, while *Section 5* discusses the main results. *Section 6* is devoted to concluding remarks.

## 2. Literature review and contribution

### 2.1 Innovation, firm heterogeneity, and skills

A large body of economic literature documents substantial heterogeneity in firms' propensity to innovate, both across and within sectors (Gil and Figueiredo, 2013). Early contributions highlight the role of sectoral patterns and technological regimes in shaping innovative behaviour (Pavitt, 1984; Bogliacino and Pianta, 2016), showing that manufacturing firms are on average more innovative than service firms, with significant variation across industries (Cainelli et al., 2006). Firm size is another key determinant: larger firms benefit from greater access to financial resources and can diversify innovation projects, reducing overall risk (Cohen et al., 1987; Kleinknecht, 1989). In addition, inter-firm relationships and knowledge spillovers play a crucial role, as localized externalities and open innovation practices facilitate access to external knowledge, often triggering innovation processes (Glaeser et al., 1992; Chesbrough, 2004). Finally, the institutional context, encompassing education systems, labour market institutions, and innovation policies, strongly affects both the likelihood and the type of innovation (Lundvall, 1992; Nelson, 1993; Datta et al., 2019). While these contributions identify several correlates of innovative performance, they converge on a common insight: persistent differences in firms' innovative outcomes are closely linked to heterogeneity in skill endowments and in the processes through which skills are formed, combined, and deployed within firms. Understanding the sources of innovation heterogeneity therefore requires a closer examination of how skills are defined and accumulated at the micro level.

## 2.2 Human capital approach to skill formation

In mainstream economic theory, skill formation is primarily analysed through the lens of human capital accumulation. Following the widely adopted Anglo-Saxon approach, a task is defined as a discrete unit of work activity contributing to economic output, while a skill is the ability to perform such a task (Rodrigues et al., 2021). Tasks and skills do not map one-to-one, as task performance typically relies on combinations of multiple skills. From this perspective, firms can adjust their skill endowment either internally, through training, or externally, through the recruitment of workers from the labour market. When training takes place within firms, it is based on an exchange between employers and employees: firms invest in workers' skills to raise productivity, while workers receive higher wages in return.

The modern economic analysis of skill formation originates with Becker (1964), who distinguishes between general and firm-specific training. General training produces skills that are transferable across firms and industries, whereas firm-specific training generates skills that are valuable only within the firm where training takes place. Firms have no incentive to finance general training in condition of perfect competition, while this result does not hold if labour market imperfections, such as wage compression or imperfect information, are introduced (Acemoglu, 1997; Acemoglu and Pischke, 1998 and 1999). Other contributions further refine this framework by emphasizing the combinatorial nature of skills. Lazear (2009), for instance, conceptualizes workers as endowed with bundles of interacting skills whose combined value may be firm-specific, even when individual skills are general. The human capital approach attempts to explain the internal process of skill development as being caused by an expected increase in the performance of individual workers, without implying any structural change. This framework has recently been extended by the task-based approach proposed by Acemoglu et al. (2011), which conceives jobs as bundles of tasks and skills. Within this approach, exogenous technological change reallocates tasks across low-, medium-, and high-skilled workers, primarily by automating routine tasks.

Together, these approaches provide a coherent account of how individual skills are formed and allocated in response to technological change. However, they tend to overlook the institutional and organizational environments in which skills are actually developed and utilized.[2]

### 2.3 Institutions, skill development, and dynamic capabilities

Addressing these limitations, the institutionalist literature emphasizes that skill formation is deeply embedded in the organizational and institutional context in which firms operate. In particular, a large body of work emphasizes that skill development is embedded in firms' internal labour markets, where job boundaries, wage-setting, and career paths are administratively regulated rather than determined through market-based interactions (Doeringer and Piore, 1971; Williamson, 1975, 1991; Baker et al., 1994). Access to internal labour markets is typically limited to specific entry positions, while internal mobility follows structured career paths, along which skills are progressively developed through on-the-job training. The way skills and tasks are grouped within firms thus reflects institutional arrangements rather than purely technological considerations.

At the same time, not all skills are developed internally. Workers with medium to high levels of qualification often rely on occupational labour markets, where competencies are acquired and validated externally through formal education, certification systems, or professional communities (Marsden, 1999). In practice, most jobs combine internally developed skills with externally acquired and certified competencies, which are subsequently adapted to firm-specific needs. The balance between internal skill development and external skill acquisition is thus shaped by institutional configurations governing education systems, training regimes, and labour market regulation (Christenko, 2020; Estevez-Abe et al., 2001). These configurations have important implications for the portability of skills and for firms' ability to adjust their workforce in response to structural change.

---

[2] This setting is also heavily influenced by the institutional framework. Indeed, the definition of skill is quite controversial and deserves some consideration as the definition adopted affects the analysis of the process itself of skill development in firms. As outlined by Toner (2011) and, particularly, Clarke and Winch (2006), there are two contending definitions of skill: the Anglo-Saxon and the German. The Anglo-Saxon notion of skill emphasises the techno-productive dimension: a skill consists of the ability to perform a set of coordinated tasks inside a productive environment. A rather different definition is that of Greene (2013), who refers to skills as personal qualities with three features: productive, expandable through training, and acquired through social interactions such as education, on-the-job training, and coaching. The German notion of skill emphasises the social dimension of productive activity and its connection to general education. The emphasis is less on the tasks to be performed than on the ability to apply what has been learned in theory through the practice of autonomous thinking and problem-solving. This skill concept is inherently linked to the formal recognition of qualifications attained in an educational or training institution. This last aspect is notably absent from the Anglo-Saxon notion of skill.

Evolutionary economics builds on this institutional perspective by explicitly linking skill development to organizational learning and innovation. Teece et al. (1997) introduce the notion of dynamic capabilities, defined as firms' ability to integrate, build, and reconfigure internal and external competencies in response to changing technological and market conditions. From this viewpoint, not all skill formation contributes equally to innovation; only those embedded in cumulative learning processes enable firms to sustain competitive advantage (Argyris and Schön, 1978). Closely related is the concept of absorptive capacity, originally defined as the firm's ability to recognize, assimilate, and apply external knowledge for the development of technological innovations (Cohen and Levinthal, 1989). Absorptive capacity depends critically on firms' skill endowments and on how skills are organized, combined, and deployed within the firm. Zahra and George (2002) further refine this notion by emphasizing the transformation and exploitation of knowledge, arguing that firms endowed with stronger capabilities along these dimensions are more likely to achieve innovation-driven growth.

Recent contributions have adopted an explicitly evolutionary perspective, bringing empirical support to the mechanisms discussed in the dynamic capabilities and absorptive capacity literature. These approaches particularly emphasise the importance of internal training and the recruitment of skilled employees for the processes of skill development required for managing innovation. Cirillo et al. (2023) and Cucculelli et al. (2025) show that employees' skill endowments affect firms' adoption of Industry 4.0 technologies, while Mason et al. (2020) links skills to innovative output within an absorptive capacity framework. Importantly, this line of research moves beyond a unidirectional view of skills as an input to innovation, as some studies highlight the endogeneity of the skills-innovation nexus, showing that innovation feeds back into upskilling practices and learning processes within firms (Pedota et al., 2023).

**Table 1** summarizes the main empirical contributions reviewed in this section, highlighting the theoretical perspective, topic, data sources, and key findings.

**Table 1 -** Summary of empirical literature on skills, innovation, and firm performance

| Study | Theoretical perspective | Focus | Data | Main findings |
|---|---|---|---|---|
| Dearden et al. (2006) | Human capital | Training and productivity | UK firms | Work-related training increases productivity (value added) |
| Zwick (2006) | Human capital | Training and productivity | German firms | Positive correlation between training and productivity (value added) |
| Bauernschuster et al. (2009) | Human capital/Innovation | Training and innovation | German firms | Continuous training increases a firm's propensity to innovate (not radical innovation) |
| Dostie (2018) | Human capital/Innovation | Training types and innovation | Canadian firms | On-the-job and classroom training have a positive impact on workplace-level innovation performance. |
| González et al. (2016) | Human capital/Innovation | Training and R&D | Spanish firms | Training complements R&D in fostering innovation |
| Mason et al. (2020) | Evolutionary | Skills and absorptive capacity | European firms | Skills enhance innovative output via absorptive capacity |
| Calvino et al. (2022) | Evolutionary / Task-based | Automation and tasks | OECD countries | Automation reshapes task and skill demand |
| Cirillo et al. (2023) | Evolutionary | Skills and Industry 4.0 | Italian firms | Skill endowment increases probability of adopting I4.0 technologies |
| Caravella et al. (2023) | Evolutionary | Digital skills and innovation | EU firms | Digital skills support technological upgrading |
| Pedota et al. (2023) | Evolutionary | Innovation and skills | Italian firms | Technology adoption feeds back into upskilling strategies |

| Cucculelli et al. (2025) | Evolutionary | Training costs | Italian firms | Training costs and skills shape advanced technology investments |

## 3. A synthesis between the neoclassical and institutionalist-evolutionary approaches

Two strands of literature suggest that firm-level heterogeneity in innovation cannot be fully explained by sectoral characteristics, firm size, or technological opportunities alone. It reflects systematic differences in how skills are formed, organized, and embedded within firms, processes that are themselves shaped by institutional environments and organizational structures.

To integrate the evolutionary/institutional perspective with the standard neoclassical one, an adjustment cost function is introduced. We assume that any innovation, whether in terms of products or processes, will require changes to the firm's employment structure. These changes may result from various internal adjustments within the firm. As a general rule we assume that, when a firm innovates, it can decide to either train current employees or hire additional employees with the necessary skills. Besides, both strategies can be adopted by the firm.

An adjustment cost function is introduced following Hamermesh (1995). According to this approach, lumpy costs arise whenever the firm hires or adjusts employment. Let us assume that the firm incurs labour adjustment costs whenever it implements a technological or organisational innovation. Three scenarios can be conceived in this respect:

a) Adjustment costs arise because the innovation requires training for some employees ($CT$).

b) Adjustment costs arise because the innovation requires skills that are not present within the firm, so new employees with the necessary skills are recruited ($CH$).

c) Adjustment costs arise because both training current employees and recruiting new employees with the necessary skills are required.

Consequently, the adjustment cost function consists of two components. First, we indicate with $KL$ the costs associated with training current employees, and with $T$ the number of trained employees. Second, there are the costs associated with hiring new employees who possess the necessary skills to manage the innovation introduced by the firm (denoted by *KH*). $H_t$ is the number of newly recruited employees who have the required skills. $L^*$ is the level of employment at which profits (*Π*) are maximised, net of adjustment costs, and $L_{t-1}$ represents the level of employment prior to the introduction of technological or organisational innovations, the adjustment cost function is given by:

$$CT = KL \; if \; T > 0 \; and \; H_t = 0$$

$$CH = KH \; if \; H_t > 0 \; and \; T = 0$$

$$CT = KL + KH \; if \; H_t > 0 \; and \; T > 0$$

Accordingly, if

$$KH + \Pi\left(L^*\right) - \Pi\left(L_{t-1}\right) > KL + \Pi\left(L^*\right) - \Pi\left(L_{t-1}\right)$$

then we have that:

$$H_t = 0$$

In this case, the technological innovation results in the training of present employees. However, when we have that:

$$KH + KL > KL + \Pi\left(L^*\right) - \Pi\left(L_{t-1}\right)$$

which implies:

$$KH + \Pi\left(L^*\right) - \Pi\left(L_{t-1}\right) < KL + \Pi\left(L^*\right) - \Pi\left(L_{t-1}\right)$$

technological innovation results in the recruitment of new employees with suitable skills:

$$H_t > 0 \; and \; T = 0$$

Finally, if:

$$KH + KL > KH + \Pi\left(L^*\right) - \Pi\left(L_{t-1}\right)$$

which gives

$$KH + KL < KL + \Pi\left(L^*\right) - \Pi\left(L_{t-1}\right)$$

and

$$KH + KL < KH + \Pi\left(L^*\right) - \Pi\left(L_{t-1}\right)$$

Firms will pursue a strategy of hiring new worker and training exiting ones:

$$H_t > 0 \; and \; T > 0.$$

This cost function suggests that technological innovations are associated with the implementation of internal training practices and the recruitment of skilled workers.

Adopting Becker's analytical framework, one could argue that specific training is associated with internal training practices, whereas acquiring general skills is correlated with the recruitment of individuals with these skills. This is because employers are unwilling to finance general training due to the hold-up argument. However, if one considers that performing a job requires a combination of various skills (Lazear, 2009; Guidetti and Mazzanti, 2007) and that firms differ in the way they combine these skills, then any general skill becomes specific as a result of this combination. The use of each general skill is tailored according to the firm's specific requirements. Consequently, the use of a general skill becomes firm-specific, which may make employees reluctant to invest in it. Therefore, any form of innovation that requires the recruitment of employees with general skills will also require at least a minimum amount of internal learning. However, this does not imply that the recruitment of highly skilled employees should necessarily be followed by formal internal training. Learning can also occur informally through on-the-job training and various unstructured, uncodified and unmeasurable processes of transmitting tacit knowledge. These considerations lead us to conclude that internal learning and training activities can only be captured through proxies, and that direct measures of training activities do not provide a comprehensive account of all learning and training activities within firms.

### 3.1 The research questions

It should be clear from our economic literature review that two alternative approaches can be identified. On the one hand, we have the human capital approach, which emphasises the role of the market in implementing training practices to boost firm innovation performance. In this approach, training does not lead to structural change within the firm but is merely a tool for optimising employees' working time. On the other hand, we have the institutionalist/evolutionary approach, in which skills development within firms is driven by technology or specific institutional settings. In this second approach, the role of the interactions between labour demand and supply is barely mentioned or even neglected. Training is a tool for adapting the workforce's characteristics to structural change.

To bring these two approaches together, the cost function developed in Section 3 highlights that implementing innovative practices in firms, such as process, product and circular innovation, necessitates existing employees training or the recruitment of lacking skills from new workers, or a combination of both. This framework motivates the following research questions:

**RQ1** Are human resource practices, namely internal training and external recruitment, associated with innovation adoption?

**RQ2** Do different forms of innovation require specific skill-development approaches, or are both internal training and external recruitment required across all innovation types?

## 4. Context, data, and methodology

### 4.1 The Italian context

The Italian macroeconomic framework is characterised by a diversified economy, with a strong industrial base in the North focused on high-quality manufacturing, mainly machinery, vehicles, and fashion. Despite being the third-largest economy in the Eurozone, Italy lags behind in innovation and employees training due to structural inefficiencies (European Innovation Scoreboard, 2023; ISTAT, 2023). According to the European Innovation Scoreboard, Italy is classified as a "moderate innovator", with 51% of firms deemed innovative, compared to the EU average of 60% (European Innovation Scoreboard, 2023). Between 2013 and 2023, Italy's R&D expenditure increased marginally, from 1.29% to 1.31% of GDP, remaining below the EU average of 2.22% (Eurostat, 2025[3]). However, patent performance has improved, with Italian applications to the European Patent Office rising 3.8% in 2023, ranking Italy fifth in the EU and eleventh globally (European Patent Office, 2023). Additionally, although Italy has enhanced its connectivity and digital tool adoption (e.g., 5G and cloud computing), it lags in advanced technologies like AI and big data, ranking 18th in the EU in 2022.

Regarding adult participation in training activities, Italy lags significantly behind other OECD countries. In 2019, only two in ten adults engaged in job-related training, and just 60.2% of firms with at least 10 employees provided training for their workers, below the OECD European average of 76% (OECD, 2019). During the 2020-2021 period, Italy's adult training participation dropped sharply to 7.2%, returning to 2015 levels due to pandemic restrictions. Despite narrowing the gap with the European average of 9.2%, its ranking remained low at 16th in Europe (ANPAL and INAP, 2024). The total number of hours dedicated to training in Italy is also lower with respect to the EU-27 average (133 hours versus 144) due to the fewer hours generally allocated to formal education (ISTAT, 2024). The reasons behind these gaps in training and education are multifaceted. However, they are primarily driven by several structural factors, including regional and local disparities between the North and the South, demographic ageing, and limited access to training opportunities for vulnerable groups, such as individuals with low educational attainment or low professional qualifications, people over 50, and people receiving social benefits (ReferNet Italy and Cedefop, 2024). Furthermore, Italy lacks significant investment in public and private training (Enzima12, 2023).

These data highlight and confirm that a key issue in Italy is the gap between large companies and small to medium-sized enterprises, with larger firms investing more in training, particularly in more industrialised regions, thus highlighting a structural kind of issue. In contrast, small enterprises, which comprise a large part of the economy, often face financial constraints and access to limited resources (Sistema Informativo Excelsior, 2023). Employee participation in training also varies across sectors, with the highest rates in utilities (64.2%), construction (57.4%), business services (55.7%), personal services (54.1%), and manufacturing (52.6%). Training activity is particularly strong in the chemical and pharmaceutical industries (67.2%), business support services (consulting: 61.2%), financial and insurance services (74.5%), and healthcare (69.6%). The tourism sector has also grown (39.1%), though it remains still undersized (Enzima12, 2023).

As innovation accelerates and the labour market and socioeconomic landscape rapidly change, continuing education systems are challenged to adapt with greater flexibility and efficiency to the demands of digital and green skills. For these reasons, EU policies and national initiatives emphasise training as a critical tool for providing workers with the skills needed for digital and green transitions while boosting workforce competitiveness (ANPAL and INAP, 2024).[4]

### *4.2 Dataset and questionnaire*

The data collection was based on an original dataset of small and medium-sized Italian enterprises observed during 2017-2018 and 2019-2020. The data were collected through two national surveys conducted in 2020 and 2021. Both surveys were administered via a Computer Assisted Web Interview (CAWI) by the company Izi s.p.a. A structured questionnaire, consisting of four main sections, was used to gather firm-level information on the role of innovation practices.

The questionnaire has been developed based on information from existing official EU sources, including Community Innovation Survey and Eurobarometer surveys. It is structured into four main sections: 1) Firms' characteristics; 2) Innovation and investments; 3) Circular economy; 4) Training and industrial relations[5].

The initial segment aims to gather general information about the firms, including their location, industry classification, respondent details, turnover over two years, company age, export levels, the number of employees over two years, and educational background. The second part assesses innovation activities, distinguishing between process and product innovation. Additionally, it delves into the firms' capacity for investment in R&D dedicated to reducing the environmental impacts during production and the adoption of patents. The third segment is specifically centred on the adoption of circular economy innovation practices. This section also examines potential drivers for adopting innovation, comparing market-based and non-market-based instruments. The last section evaluates the significance of environmentally friendly high-performance practices, such as organisational training and reskilling initiatives designed to manage the shift to a circular economy. It also considers the role of industrial relations in the adoption of circular economy innovation.

---

[4] For instance, the Italian National Recovery and Resilience Plan (PNRR) has recently included a range of initiatives designed to increase participation in training programs, improve workforce adaptability, and support vulnerable groups in accessing education and reskilling opportunities.

[5] Several parts of the original questionnaire that refer to our main variables of interest are reported in Table A3 in the Appendix.

The sample includes 4,565 firms in the first wave and 4,649 in the second, with 2,305 firms responding to both surveys. The sample is stratified by geographical location (macro area, ISTAT), sector (technological intensity, NACE[6]), and size (10-49 employees; 50-249 employees; >250 employees).

The original balanced dataset, counting 4,610 observations, was subsequently merged with balance sheet data from the Bureau van Dijk AIDA dataset, resulting in the loss of 980 observations. Additionally, data on employed individuals per Province, number of inhabitants, workforce, and manufacturing employees for the years 2018, 2019, and 2020 were incorporated from Italian National Institute of Statistics data. Subsequently, data on high school presence per province were added, relying on the online database of Comuni & Città [7]. Finally, firms operating in sectors other than manufacturing were excluded, resulting in the loss of 52 observations. The final sample comprises 3,578 observations, corresponding to 1,789 per year.

### 4.3 Measuring innovation adoption and training

The dependent variables are intended to capture the introduction of product innovation (*ProdInno*), process innovation (*ProcInno*), circular economy innovation (*Cei*), and total innovation (*InnoTot*).

---

[6] Eurostat classification https://ec.europa.eu/eurostat/statistics-explained/index.php?title=Glossary:High-tech_classification _of_manufacturing_industries

[7] https://www.comuniecitta.it/scuole-italiane

*ProdInno* is a dummy variable that equals 1 for firms that have introduced a technologically new (or significantly improved) product or service compared to those previously available in terms of technical and functional characteristics and performance (40% in the biennium 2017-2018 and 35% in the biennium 2019-2020). *ProcInno* is a dummy variable that assigns a value of 1 to those firms that have introduced new (or significantly improved) methods compared to those previously adopted by the company, considering technical and functional characteristics and performance (43% in the biennium 2017-2018 and 38% in the biennium 2019-2020). Both types of innovations need not necessarily entail new-to-the-market processes, products, or services. It is sufficient that they are new to the firm that introduced them.

The adoption of *Cei* has been explored by posing a series of questions related to firms' implementation of innovations aligned with the pursuit of circular economy strategies. It includes innovations aimed at (i) minimizing water usage, (ii) minimizing raw materials usage, (iii) increasing renewable energy usage, (iv) minimizing fossil-based energy use, (v) minimizing waste generation, (vi) reusing waste within own firm productive process, (vii) delivering waste to other companies to be reused in their productive process, (viii) redesigning products to minimize the use of materials, (ix) redesigning product to enhance their recyclability, (x) changing the production process to reduce greenhouse gas emissions. A company has adopted a *Cei* if it has introduced at least one of the above-mentioned innovations. It results that *Cei* is a dummy variable that takes a value of 1 if the firm has adopted one or more circular-oriented innovative practices. In our sample, *Cei* equals 1 for 44.77% of firms between 2017- 2018 and 38.68% between 2019-2020.

Finally, *InnoTot*, indicate whether the firm has adopted at least one of the three innovation practices. This variable is equal to 1 if the firm has adopted at least one of the three practices and 0 otherwise.

Our primary regressors evaluate the company's sensitivity level towards the training of its employees. The respondents were asked to indicate, for the two-year periods 2017-2018 and 2019-2020, the percentage of employees who are annually involved in training activities (mentoring, with structured programs, or training courses). *Trained Employees* is a continuous variable calculated as the percentage of employees engaged in training activities. *Stem* is a binary variable with a value of 1 if the respondent declared that scientific, technological/engineering, mathematical, and informatics skills play an 'Absolutely relevant' or 'Very relevant' role in the organisation and 0 otherwise.

The variable *Trained Employees* can be linked to the human capital approach, as it can be interpreted as an indicator of the efforts made by both employers and employees to improve the workforce's skills through general and specific training. *Stem* can also be associated with strategies aimed at strengthening firms' absorptive capacity by upgrading the endowment of general skills. In this case, the *Stem* variable would proxy for a more general and formalised kind of training (obtained at school), which also depends on the different institutional contexts within which firms are embedded.

These two variables deserve special attention as they play a special role in our estimation strategy. The variable *Trained Employees* can be employed as an exogenous control variable to gauge the firm's endeavours in overseeing internal training programs for employees, irrespective of any innovation strategy. Nevertheless, it is possible to endogenize the variable by introducing the hypothesis that the decision to train employees interacts with the firm's innovation policies. *Stem* can be used either as an exogenous control variable to control for firms' recruitment policies, or as an endogenous treatment to address how recruitment policies interact with internal training practices.

Our third main variable is *Labour Cost*, which is calculated as the cost of labour divided by the number of employees in 2018, in logarithms.

The other explanatory variables encompass factors recognised in existing literature as drivers of innovative activities. They include *%Graduates* representing the employees' educational level, as a proxy of knowledge stock, which is measured by the share of employees with a degree of tertiary education. *High School by Employee* is a continuous variable indicating the presence of high schools divided by workers in the manufacturing sector in 2018, in logarithms. *R&D* is a dummy variable with a value of 1 if the firm undertakes R&D investments, and *Green R&D is* a dummy variable with a value of 1 if the firm undertakes R&D investments to reduce environmental impacts. Other firms' characteristics include *Ebitda,* which is the ratio of EBITDA to the number of employees in 2018, *Age*, which is a continuous variable indicating firms' age. *Size* is a continuous variable indicating the number of employees. *Exporter* is a dummy variable that takes the value of 1 for exporting firms. *Industrial Group* is a dummy variable with a value of 1 if the firm belongs to a group of firms. Finally, *Trade Unions* is a dummy variable that assigns a value of 1 if union representatives exist in the firm.

We subsequently included the following control variables. *Scale Intensive*, *Science-Based*, *Specialised Suppliers*, *Supplier Dominated,* are four binary variables constructed following the Pavitt methodology in order to measure the technological regimes of the firms' sector. We also consider two work-related measures with objective measurement scales as instruments. *Candidate Interviews* is a categorical variable indicating how frequently job candidates within the respondent's firm undergo structured interviews (ranging from 1 to 5, where 1 is "Never" and 5 is "Very Often"). *Employee Training Reimbursement* is a categorical variable indicating how frequently employees receive reimbursement for external training courses (this variable ranges from 1 to 5, where 1 is "Never", and 5 is "Very Often").

**Table 2** reports the summary statistics of the variables included in the model[8].

**Table 2** – Descriptive statistics

| Variable | Description | Mean | Std. dev. | Min | Max |
|---|---|---|---|---|---|
| *ProdInno* | Introduction of Product Innovation | .37758 | .48485 | 0 | 1 |
| *ProcInno* | Introduction of Process Innovation | .40245 | .49046 | 0 | 1 |
| *Cei* | Introduction of Circular Innovation | .41727 | .49317 | 0 | 1 |

---

[8] Given the number of covariates, we check for collinearity first through bivariate correlation matrices (see Tables A1 and A2 in Appendix) and subsequently by running an ancillary regression and computing the Varian Inflation Factor (VIF), which is equal to 1.43. Both the inspection of the correlation matrices and the meanVIF value do not rise concerns on multicollinearity.

| | | | | | |
|---|---|---|---|---|---|
| *InnoTot* | Introduction of Product Innovation or Process Innovation or Circular Innovation | .62017 | .48541 | 0 | 1 |
| *Trained Employees* | Annual percentage of employees involved in training activities | 31,086 | 35,457 | 0 | 100 |
| *Stem* | Firm with STEM competences | .43543 | .49588 | 0 | 1 |
| *% Graduates* | Share of employees with a degree who have completed tertiary education | .58724 | 2,171 | 0 | 40 |
| *High School by Employees* | Number of high schools in provinces/ number of employees in the manufacturing sector (2018) | -6,433 | .71611 | -7,474 | -3,509 |
| *R&D* | Investments in R&D | .33454 | .47189 | 0 | 1 |
| *Green R&D* | Investments in R&D aimed at reducing environmental impacts | .08671 | .28145 | 0 | 1 |
| *Unit Labour Cost* | Labor cost/number of employees in 2018 | 7,368 | .78929 | 3,618 | 10,688 |
| *Ebitda* | EBITDA / number of employees 2018 | .43704 | 1,307 | -55,587 | 23,495 |
| *Scale Intensive* | Technological regime of the firms' sectors according to Pavitt methodology | .20905 | .40669 | 0 | 1 |
| *Science-Based* | Technological regime of the firms' sectors according to Pavitt methodology | .04974 | .21745 | 0 | 1 |
| *Specialised Suppliers* | Technological regime of the firms' sectors according to Pavitt methodology | .21045 | .40768 | 0 | 1 |
| *Supplier Dominated* | Technological regime of the firms' sectors according to Pavitt methodology | .53074 | .49912 | 0 | 1 |
| *Age* | Firm age | 29.104 | 21.104 | 1 | 200 |
| *Size* | Firm dimension | 35,238 | 76,63 | 0 | 1,334 |
| *Exporter* | Exporting firms | .50810 | .50000 | 0 | 1 |
| *Industrial Group* | Firm belonging to an Industrial group | .14337 | .35050 | 0 | 1 |
| *Trade Unions* | Presence of union representatives in the firm | .24147 | .42803 | 0 | 1 |
| *Candidate Interviews* | Frequency with which job candidates undergo structured interviews. | 3,066 | 1,38 | 1 | 5 |
| *Employee Training Reimbursement* | Frequency with which employees are reimbursed for expenses related to external training courses | 3,514 | 1,571 | 1 | 5 |

## *4.4 The strategy of empirical analysis*

The theoretical framework developed in the previous sections highlights the gap between the analysis of innovation adoption practices and the internal training needs of firms, which depends on how different approaches define training activities. The human capital approach addresses this issue in terms of the optimal allocation of resources. In contrast, the institutional/evolutionary approach focuses, in particular, on the processes of introducing techno-organisational innovations in a specific institutional setting and the resulting competitive advantage of the firm (Lam and Lundvall, 2007). Consequently, human capital analysis aims at identifying the optimal distribution of employees' time between work and training activities to enhance the firm's performance. In contrast, the capabilities approach is primarily concerned with examining the impact of training on the firm's competitive advantage. Consequently, whereas in the human capital approach, the allocation of individual workers' time plays a key role in improving workers' skills and, hence, firm performance, the institutionalist/evolutionary approach suggests that skills develop as a result of the structural/organisational characteristics of the firm, thus driving the process of skills upgrading and the development of its absorptive capacity.

This paper aims to synthesise these two approaches by examining the links between the adoption of innovation practices and the implementation of strategies aimed at upgrading employees' skills through the introduction and management of internal training. As previously outlined in the dataset description, three distinct dummy variables are available as dependent variables, each indicative of the adoption of one of three forms of innovation: namely, process, product, and circular.

Hence, we estimate the impact of internal skill development on firms' propensity to adopt the (different types of) innovation.

To address our first research question, the variable *Trained Employees* is considered as an indicator of internal skills development, indicating the percentage of employees involved in internal training practices. It is important to highlight the different meanings that this variable can assume, depending on its impact on the propensity to innovate. *Trained employees* is an indicator of skills acquisition processes taking place within the firm. The implementation of training practices can engender a range of interventions that address the characteristics of employees. These may enhance skills, but they may also be designed to facilitate a match between employees and the firm. It can be concluded that implementing internal training practices does not necessarily lead to an increase in firms' absorptive capacity. An increase in absorptive capacity can only be observed if one of these variables positively affects the propensity to introduce an innovation.

On the other hand, *Stem* can indicate how firms rely on external knowledge formation by referring their recruitment policies to the education system and, thus, to general (non-specific) training. We will therefore use these two variables to detect how firms manage the acquisition of the skills they need, either internally (through a specific kind of training) or externally (through a general type of training).

Another crucial element to consider is that, in examining the relationship between training and firms' propensity to innovate, the issue of reverse causality strongly emerges. This gives rise to problems of endogeneity, resulting in inconsistent estimates of the parameters of standard models. Consequently, our estimation strategy will be based on three distinct procedures. First, we provide the results stemming from a Probit model. Four probit models are thus estimated, with the dependent variable representing a dummy variable for each of the forms of innovation outlined in the questionnaire: process innovation, product innovation, circular innovation, and total innovation. This can be considered the baseline model of our analysis.

To address the problem of endogeneity given by reverse causality, we employ a two-stage procedure utilising an instrumental variable, following the procedure outlined by Rivers and Vuong (1988) and Newey (1987) for estimating a Probit model with an endogenous explanatory variable. In this approach, Newey computes a two-stage conditional Maximum Likelihood model consisting of a structural equation, our primary interest, and a reduced equation for the endogenous variable.

We use *Candidate Interviews* and *Training Reimbursements* of employees as instruments. These two variables have been selected because they appear to be correlated with the endogenous regressor and not correlated with the model's error. The properties needed for an effective instrument are the following: a) the instrument must be correlated to the endogenous variable. If the correlation is low, we are in the presence of a weak instrument; b) the correlation between the instrument and the model's error of the estimate must be statistically insignificant: i.e., the instrument is exogenous. The first condition is relatively easy to test, whereas the exogeneity condition cannot be tested (Verbeek, 2017). In the linear case, to address point a), one must estimate the reduced form regression and then calculate the F-statistics for the significance of the instruments. As a rule of thumb, Stock and Watson (2007) claim that, in the linear case, if the F-statistic is greater than 10, there is no need to worry about instrument weakness. In a nonlinear model, such as ours, the Wald test is an equivalent test to the F-statistic, even though it is not reliable for testing instrument weakness. The exclusion restriction implies that the instrument is excluded from the main equation. As Verbeek (2017, p.154) states "the fact that scope for testing the validity of instruments is very limited indicates that researchers should pay careful attention to the justification of their instruments. The reliability of an instrument relies on argumentation, not on empirical testing". In our case, the questionnaire asks whether reimbursement for employees' training is available (*Are employees reimbursed for costs incurred when attending training programs outside the firm?*). We thus maintain that reimbursement of employees for external training and structured interviews for job applicants are procedures used routinely that are not influenced by the propensity of firms to adopt any form of innovation.

We then proceeded to the second stage of the empirical analysis by estimating a probit model with two endogenous variables. The reduced-form equation of this empirical model expresses the endogenous variable $X_{it}$ as a function of the instrumental variables and all exogenous regressors in the model. With two instruments, the reduced-form equation is written as:

$$X_{it} = \pi_1 Z_{1,it} + \pi_2 Z_{2,it} + \delta' W_{it} + \alpha_i + \lambda_t + u_{it}$$

Where $\pi_1$ and $\pi_2$ measure the effect of each instrument on the endogenous variable, $u_{it}$ is the reduced-form error term. This equation captures the variation in $X_{it}$ that is driven exclusively by the instrumental variables, $Z_{1,it}$ $and$ $Z_{2,it}$, and the exogenous controls $W_{it}$. In our case the two instruments are given by *Stem* and *Trained Employees*.

## 5. Results

### 5.1 The baseline panel probit

The results of the baseline panel model are presented in **Table 3**, which shows the marginal effects. These results show that the propensity of Italian firms to innovate is influenced differently by our primary regressors (*Trained Employees* and *Stem*). *Stem* is associated with all the forms of innovation measured in this study. Conversely, the variable *Trained Employees* is significantly correlated only with circular innovation and with process innovation. We can state that these first results lead us to positively answer RQ1, as far as circular and process innovation are concerned. On the other hand, these estimates highlight a non-significant relationship between internal training and product innovation. The combined results of our two main covariates induce us to answer to our RQ2 in these terms: although training activities and skills development are crucial for all types of innovations, it emerges that internal training seems to be more useful when circular innovations are implemented, while external recruitment seems to be always needed when innovations, irrespectively of the type, are introduced. Hence, the overall relation between training and innovation is assessed in our empirical investigation, as expected, but nuances on the specific strategies needed to upskill the workforce in presence of different innovations emerge too.

**Table 3** – Baseline Probit estimation, marginal effects

|  | *InnoTot* | *CeI* | *ProcInno* | *ProdInno* |
|---|---|---|---|---|
| *Trained Employees* | 0.000250 | 0.000389* | 0.000185 | 0.000533** |
|  | (0.000214) | (0.000217) | (0.000207) | (0.000225) |
| *Stem* | 0.134*** | 0.119*** | 0.0849*** | 0.0831*** |
|  | (0.0152) | (0.0157) | (0.0143) | (0.0166) |
| *% Graduates* | 0.00231 | 0.00339 | 0.00730* | 0.00284 |
|  | (0.00546) | (0.00397) | (0.00435) | (0.00422) |
| *High School by Employees* | -0.00171 | 0.00176 | 0.0170 | 0.00656 |
|  | (0.0175) | (0.0175) | (0.0159) | (0.0190) |
| *R&D* | 0.355*** | 0.289*** | 0.339*** | 0.160*** |
|  | (0.0181) | (0.0164) | (0.0127) | (0.0184) |
| *Green R&D* | 0.142*** | 0.0816*** | 0.0495** | 0.303*** |
|  | (0.0337) | (0.0277) | (0.0243) | (0.0308) |
| *Unit Labour Cost* | -0.0486*** | -0.0606*** | -0.0167 | -0.0546*** |
|  | (0.0134) | (0.0140) | (0.0120) | (0.0146) |
| *Ebitda* | 0.0112 | 0.0183** | -0.00121 | 0.00862 |
|  | (0.00773) | (0.00913) | (0.00375) | (0.00873) |
| *Scale Intensive* | 0.0774*** | 0.0994*** | -0.0191 | 0.111*** |
|  | (0.0257) | (0.0262) | (0.0232) | (0.0283) |
| *Science Based* | 0.00249 | -0.0204 | -0.00177 | 0.00284 |
|  | (0.0399) | (0.0395) | (0.0355) | (0.0439) |
| *Supplier Dominated* | 0.0444** | 0.0404* | -0.0496*** | 0.0649*** |
|  | (0.0213) | (0.0222) | (0.0192) | (0.0234) |
| *Age* | 0.000467 | -0.0000132 | 0.000402 | 0.00105** |
|  | (0.000415) | (0.000387) | (0.000358) | (0.000427) |
| *Size* | -0.000181 | -0.000291** | -0.0000914 | -0.000318** |
|  | (0.000144) | (0.000143) | (0.000112) | (0.000158) |
| *Exporter* | 0.0790*** | 0.0492*** | 0.136*** | 0.0612*** |
|  | (0.0166) | (0.0179) | (0.0154) | (0.0187) |
| *Industrial Group* | -0.0182 | 0.00166 | 0.0255 | -0.0185 |
|  | (0.0251) | (0.0257) | (0.0220) | (0.0279) |
| *Trade Unions* | 0.0162 | -0.00904 | -0.0296 | 0.0573*** |
|  | (0.0199) | (0.0203) | (0.0186) | (0.0211) |
| *North* | 0.0416* | 0.0199 | 0.0218 | 0.0348 |
|  | (0.0224) | (0.0233) | (0.0216) | (0.0253) |
| *South* | 0.0330 | -0.0234 | 0.0503 | 0.0599 |
|  | (0.0352) | (0.0364) | (0.0333) | (0.0380) |
| N. Obs. | 3509 | 3509 | 3509 | 3509 |
| Chi-squared | 417.4 | 384.9 | 456.5 | 324.3 |

*Notes:* Reported coefficients are marginal effects calculated at the mean of all other covariates. Robust standard errors in parentheses. *** significant at the 1 % level; ** significant at the 5 % level; * significant at the 10 % level.

Of particular importance is the negative correlation between *Unit Labour Costs* and all forms of innovation, except product innovation. As unit labour costs can be considered the opportunity cost of training, this finding is consistent with the standard human capital approach. The non-significant association between this variable and product innovation suggests that this type of innovation is less cost-related than other types of innovation.

Both types of research activity (green and general R&D) are positively and highly significantly associated with all four types of innovation. The results are in line with the extant literature concerning the capacity of R&D to spur knowledge creation and, in turn innovation (e.g., Hall and Mairesse, 2006; Charlot et al., 2015).

The propensity to export has a positive effect on all four forms of innovation, indicating the role of international competition as a demand-pull force for innovation.

Following Pavitt's taxonomy (specialised suppliers is the benchmark), the results show that science-based firms are never associated with any form of innovation. Supplier-dominated and scale-intensive firms are associated with process and circular innovation, but not product innovation. That is to say: science-based firms do not show a significant different behaviour with respect to the reference category, while supplier dominated and scale intensive firms do. Such a result might seem at odd with the usual empirical evidence that point to a relation between innovation and sector technological intensity. However, as shown by a long-standing literature on Italian firms' innovative behaviour (see as an example Santarelli and Piergiovanni, 1996), in specific sectors, belonging to the aggregated supplier dominated and scale intensive Pavitt sectors, is mostly informal (e.g. informal R&D) and tacit, mainly based on incremental changes. Such changes may also be more frequent than those supported by formal R&D projects. The same reasoning can be applied to the evidence concerning circular innovation and, in addition, we can hypothesise that firms belonging to supplier dominated and scale intensive sectors, which include many of the 'brown' sectors, have more regulatory pressure for circular innovation implementation than science based and specialised suppliers.

The impact of the other covariates differs depending on the type of innovative activity: e.g. the variable *Size* is significantly associated with process innovation (*ProcInno*), the variables *Age* and *Trade Unions* are positively associated with circular innovation.

### 5.2 The estimations with two endogenous variables

As explained in Section 3, the relationship between training and innovation appears to be vitiated by the presence of endogeneity, since, on the one side, trained employees are key for the development of innovative activities, but, on the other side, innovative activity calls for trained people to design, implement, and manage it.

Therefore, we re-run the regressions of **Table 3** by treating our primary regressors (*Trained Employees* and *Stem*) as endogenous. We thus performed a 2-stage regression. In the first stage, we regressed our two endogenous variables separately on the same set of covariates of the probit regression, plus one instrument for each of the two variables: for *Trained Employees*, we used the variable *Training Reimbursements*, and for *Stem*, we used the variable *Candidate Interviews*. To test for weak instruments, in linear models, Stock et al. (2002) proposed the F-statistics for first-stage equation. For non-linear models such as ours, the Wald test can be considered equivalent to the Fisher test for linear models.

It is reasonable to assume that these two instruments are not endogenous. Reimbursement of employees for external training and structured interviews for job applicants are routinely used procedures that are not influenced by the propensity of firms to adopt any form of innovation. On the other hand, these two practices can be expected to be positively related to the variables of *Trained Employees* and *Stem*, respectively.

As a robustness check we have also estimated a control function model (Wooldridge, 2015). This is a two-stage technique where in the first stage the endogenous variables are estimated on the basis of instruments (results are shown in Table A6, A7 and A8 of the Appendix). Then, in the second stage the residuals from these first-stage regressions are included in the main equation as control functions to make regression estimates robust to endogeneity. Wooldridge (2015) maintains that "Control function methods are also very useful in panel data applications where one must account for unobserved heterogeneity as well as endogeneity".

**Table 4** – Panel Probit estimation with two endogenous variables, marginal effects

|  | *InnoTot* | *CeI* | *ProcInno* | *ProdInno* |
|---|---|---|---|---|
| *Trained Employees* | 0.00679*** | 0.00466** | 0.00500*** | 0.00982*** |
|  | (0.00204) | (0.00212) | (0.00192) | (0.00248) |
| *Stem* | 0.649*** | 0.533*** | 0.257** | 0.605*** |
|  | (0.140) | (0.150) | (0.128) | (0.156) |
| *% Graduates* | 0.00209 | -0.00137 | 0.00799* | -0.00370 |
|  | (0.00711) | (0.00540) | (0.00481) | (0.00591) |
| *High School by Employees* | -0.0438* | -0.0283 | -0.00765 | -0.0462 |
|  | (0.0263) | (0.0254) | (0.0227) | (0.0311) |
| *R&D* | 0.220*** | 0.194*** | 0.281*** | 0.0251 |
|  | (0.0354) | (0.0333) | (0.0278) | (0.0388) |
| *Green R&D* | 0.0256 | -0.000281 | -0.0113 | 0.199*** |
|  | (0.0470) | (0.0400) | (0.0345) | (0.0485) |
| *Unit Labour Costs* | -0.0847*** | -0.0852*** | -0.0281* | -0.101*** |
| | (0.0192) | (0.0200) | (0.0166) | (0.0221) |
| *Ebitda* | 0.0101 | 0.0166 | -0.0168* | 0.00145 |
| | (0.0124) | (0.0134) | (0.0101) | (0.0156) |
| *Scale Intensive* | 0.134*** | 0.147*** | 0.0115 | 0.185*** |
| | (0.0388) | (0.0376) | (0.0326) | (0.0454) |
| *Science Based* | -0.0134 | -0.0134 | -0.00116 | 0.0254 |
| | (0.0490) | (0.0484) | (0.0453) | (0.0626) |
| *Supplier Dominated* | 0.119*** | 0.108*** | -0.0220 | 0.163*** |
| | (0.0361) | (0.0352) | (0.0303) | (0.0423) |
| *Age* | 0.000592 | -0.000102 | 0.000381 | 0.00112* |
| | (0.000574) | (0.000512) | (0.000455) | (0.000657) |
| *Size* | -0.000588*** | -0.000589*** | -0.000168 | -0.000833*** |
| | (0.000178) | (0.000208) | (0.000154) | (0.000178) |
| *Exporter* | 0.0298 | 0.0212 | 0.129*** | 0.0255 |
| | (0.0276) | (0.0277) | (0.0239) | (0.0319) |
| *Industrial Group* | -0.0375 | 0.00958 | 0.0112 | -0.0518 |
| | (0.0348) | (0.0329) | (0.0282) | (0.0411) |
| *Trade Unions* | -0.0109 | -0.0336 | -0.0365 | 0.0469 |
| | (0.0297) | (0.0273) | (0.0252) | (0.0346) |
| *North* | -0.0350 | -0.0494 | -0.0184 | -0.0486 |
| | (0.0343) | (0.0337) | (0.0307) | (0.0417) |
| *South* | 0.0715 | -0.00311 | 0.0696 | 0.0801 |
| | (0.0509) | (0.0476) | (0.0436) | (0.0609) |
| *Employee training reimbursement* | 0 | 0 | 0 | 0 |
| | (.) | (.) | (.) | (.) |
| *Candidate Interviews* | 0 | 0 | 0 | 0 |
| | (.) | (.) | (.) | (.) |
| N. Obs. | 2636 | 2636 | 2636 | 2636 |
| Log Likelihood | -15883.2 | -16126.1 | -15872.3 | -16131.4 |
| Chi-squared | 883.7 | 505.2 | 425.6 | 789.6 |

*Notes:* Reported coefficients are marginal effects calculated at the mean of all other covariates. Robust standard errors in parentheses. *** significant at the 1 % level; ** significant at the 5 % level; * significant at the 10 % level.

The second-stage estimations (**Table 4**) reinforce our previous results (First stage estimation is reported in Table A4 and A5 in Appendix). Indeed, both endogenous variables (*Trained Employees* and *Stem*) acquire significance with all forms of innovation. Again, we can affirmatively answer to RQ1, while the specificities in skill development approach, that partially emerged in the baseline regression, fade out here in terms of coefficient significance, but remains in terms of coefficient magnitude: internal training seems to be more relevant for circular innovations than for other types of innovations. Also in this case, *Unit Labour Costs* confirm their role as they are significantly correlated to our innovation variables, with the smaller impact on product innovation. General *R&D* is associated to product and process innovation, but not to circular innovation. However, *Green R&D* is only significantly correlated with circular innovation. Product innovation is significantly linked to the propensity to export, but not the other innovation types as in the baseline model. All the other covariates show results consistent with those obtained in the estimation of the baseline model.

## 6. Conclusions

The aim of this paper is threefold. First, it provides an in-depth survey of the notion of skill and the process of skill development within two different approaches to economic analysis: the neoclassical approach and the institutionalist/evolutionary one. Second, it attempts to integrate these two approaches within a labour adjustment cost function. Third, it estimates the relationship between skill development within firms, whether through internal training or external skill acquisition, and the introduction of process, product, and circular innovations.

The cost function constructed in the paper represents a bridge between the human capital approach, which emphasises formal, recognised and measurable training (both general and specific), and the evolutionary/institutionalist approach, which complements this vision of training with forms of learning that are difficult to quantify, such as learning by doing, or training aimed at transmitting tacit knowledge.

Due to the difficulties involved in assessing the amount of training activities, accounting for labour adjustment costs seems rather complicated. However, the labour adjustment cost outlined in Section 3 enables the identification of possible sources of labour cost adjustment in the event of introducing innovative activities. Essentially, labour adjustment costs arising from policies designed to facilitate the introduction of innovation in its various forms can be approximated using a set of indicators, such as *Trained Employees*, *Stem*, and *Labour Cost*.

Our empirical evidence shows a significant association between our main covariates (*Stem*, *Trained Employees* and *Labour Cost*) and the propensity to adopt the three forms of innovation. *Stem* can be considered an indicator of the propensity to recruit skilled employees; *Trained Employees* is a proxy for the propensity to provide internal training to current employees; and *Labour Cost* represent the opportunity cost of training practices in firms and a proxy of the labour market conditions. All three of these variables are pivotal components of the labour adjustment costs discussed in Section 2.4.

In conclusion, and looking back to our initial research questions, firms' absorptive capacity, understood as the ability to introduce and manage all three types of innovation, is positively related to the need to recruit the required skills from the external labour market. Additionally, training within firms positively affects the likelihood of adopting any of the considered innovations. However, its impact is less pronounced than that resulting from the recruitment of skills from the external labour market.

**Appendix**

**Table A1** – Correlation matrix (Pearson correlation)

|  | Trained Employees | %Graduates | Labour Cost | Ebitda | High School by Employees | Age | Size |
|---|---|---|---|---|---|---|---|
| Trained Employees | 1 | | | | | | |
| %Graduates | 0.0731 | 1 | | | | | |
| Unit Labour Costs | -0.0128 | -0.1633 | 1 | | | | |
| Ebitda | 0.0257 | -0.047 | 0.1868 | 1 | | | |
| High School by Employees | 0.0419 | -0.0518 | -0.0801 | 0.006 | 1 | | |
| Age | 0.0344 | 0.0907 | -0.0582 | -0.0144 | -0.1512 | 1 | |
| Size | 0.0622 | 0.14 | -0.606 | -0.0776 | -0.0261 | 0.1463 | 1 |

**Table A2** – Correlation matrix (Tetrachoric correlation)

|  | Exporter | Group | R&D | Green R&D | Trade Unions |
|---|---|---|---|---|---|
| Exporter | 1 | | | | |
| Group | 0.3096 | 1 | | | |
| R&D | 0.5078 | 0.2877 | 1 | | |
| Green R&D | 0.2264 | 0.2594 | 0.4964 | 1 | |
| Trade Unions | 0.255 | 0.3734 | 0.1862 | 0.1882 | 1 |

**Table A3** – Questions related to the main variables

In the two-year period 2017-2018 / 2019-2020, you introduced:

|  | No | Yes, new only for the firm | Yes, new also for the reference sector/market. |
|---|---|---|---|
| Product Innovation | ○ | ○ | ○ |
| Process Innovation | ○ | ○ | ○ |

In the period 2017-2018 / 2019-2020, did the company introduce innovations aimed at achieving one of the following objectives related to circular economy?

|  | Yes | No |
|---|---|---|
| Minimizing water usage | ○ | ○ |
| Minimizing raw materials usage | ○ | ○ |
| Use of energy generated from renewable sources | ○ | ○ |
| Minimizing fossil-based energy use | ○ | ○ |
| Reduction of waste generated (per unit of product output) | ○ | ○ |
| Reuse of waste within the firm's own production cycle | ○ | ○ |
| Delivering waste to other companies to be reused in their productive process | ○ | ○ |
| Product design changes to minimize the use of raw materials (including energy) | ○ | ○ |
| Product design changes to maximize recyclability | ○ | ○ |
| Changes to the production process to reduce greenhouse gas emissions | ○ | ○ |

Indicate, for the two-year period 2017-2018 / 2019-2020, the percentage of employees who are involved annually in training activities (on-the-job training, structured programs, or training courses)

*123*

Indicate the level of importance that the following skills had in your organization during the two-year period 2017-2018 / 2019-2020:

| | 1. Not at all relevant | 2. Slightly relevant | 3. Moderately relevant | 4. Very Relevant | 5. Absolutely Relevant |
|---|---|---|---|---|---|
| Scientific | ○ | ○ | ○ | ○ | ○ |
| Technological / Engineering | ○ | ○ | ○ | ○ | ○ |
| Mathematical | ○ | ○ | ○ | ○ | ○ |
| Informatics/Programming | ○ | ○ | ○ | ○ | ○ |

Indicate the percentage of employees with a university degree (enter zero if there are none):

| | 2017 | 2018 | 2019 | 2020 |
|---|---|---|---|---|
| University graduates | ○ | ○ | ○ | ○ |

Did the company invest in Research and Development during 2017-2018 / 2019-2020?

- o  Yes
- o  No

Please indicate if R&D investments were made to reduce environmental impacts.

- o  Yes
- o  No

Please indicate the age of the firm

*123*

| | 2017 | 2018 | 2019 | 2020 |
|---|---|---|---|---|
| Number of employees | ○ | ○ | ○ | ○ |

Is the firm an exporter?

o   Yes
o   No

Is the firm part of an industrial group?

o   Yes
o   No

Does the company have trade union representation (RSU)

o   Yes
o   No

| | 1. Never | 2. Rarely | 3. Sometimes | 4. Often | 5. Very Often |
|---|---|---|---|---|---|
| Job candidates undergo structured interviews (job-related questions, same for all candidates, with objective rating scales) | ○ | ○ | ○ | ○ | ○ |
| Employees are reimbursed for costs of external training courses | ○ | ○ | ○ | ○ | ○ |

**Table A4** – First stage – Endogenous: Trained Employees

| | InnoTot | ProcInno | ProdInno | CeI |
|---|---|---|---|---|
| Training Reimbursements | 3.009*** | 3.001*** | 3.012*** | 3.010*** |
| | (0.456) | (0.457) | (0.456) | (0.455) |
| Unit Labour Cost | 2.864* | 2.877* | 2.849* | 2.858* |
| | (-1.214) | (-1.214) | (-1.213) | (-1.212) |
| Scale Intensive | -0.996 | -1.014 | -0.991 | -1.005 |
| | (-2.365) | (-2.366) | (-2.367) | (-2.366) |
| Science Based | -3.503 | -3.486 | -3.512 | -3.501 |
| | (-3.564) | (-3.560) | (-3.564) | (-3.565) |
| Supplier Dominated | -2.735 | -2.752 | -2.730 | -2.730 |
| | (-2.072) | (-2.071) | (-2.072) | (-2.072) |
| Size | 0.031* | 0.031* | 0.031* | 0.031* |
| | (0.012) | (0.012) | (0.012) | (0.012) |
| Exporter | -2.027 | -2.015 | -2.025 | -2.023 |
| | (-1.680) | (-1.680) | (-1.679) | (-1.680) |
| Industrial Group | 3.023 | 3.032 | 3.011 | 3.015 |
| | (-2.246) | (-2.245) | (-2.246) | (-2.245) |
| R&D | 5.136** | 5.073** | 5.212*** | 5.163** |
| | (-1.583) | (-1.576) | (-1.582) | (-1.582) |
| Green R&D | 5.293* | 5.289* | 5.313* | 5.280* |
| | (-2.268) | (-2.268) | (-2.267) | (-2.278) |
| Editda | 1.563* | 1.522* | 1.573* | 1.568* |
| | (0.721) | (0.729) | (0.721) | (0.720) |
| Trade Unions | -0.045 | -0.056 | -0.068 | -0.047 |
| | (-1.865) | (-1.863) | (-1.865) | (-1.864) |
| High School by Employees | 5.547*** | 5.541*** | 5.548*** | 5.548*** |
| | (-1.570) | (-1.570) | (-1.570) | (-1.570) |
| % Graduates | 0.195 | 0.193 | 0.196 | 0.196 |
| | (0.347) | (0.347) | (0.347) | (0.348) |
| Age | 0.019 | 0.019 | 0.019 | 0.018 |
| | (0.040) | (0.040) | (0.040) | (0.040) |
| North | 4.727* | 4.723* | 4.733* | 4.729* |
| | (-2.085) | (-2.086) | (-2.086) | (-2.086) |
| South | 2.952 | 2.961 | 2.964 | 2.956 |
| | (-3.456) | (-3.456) | (-3.457) | (-3.456) |
| Constant | 31.620* | 31.534* | 31.695* | 31.650* |
| | (14.107) | (14.108) | (14.103) | (14.101) |
| N. Obs. | 2.636 | 2.636 | 2.636 | 2.636 |

*Notes:* Reported coefficients are the results of the first stage linear estimation for the endogenous variable Trained Employees. Robust standard errors in parentheses. *** significant at the 1 % level; ** significant at the 5 % level; * significant at the 10 % level.

**Table A5** – First stage – Endogenous: STEM

| | *InnoTot* | *ProcInno* | *ProdInno* | *CeI* |
|---|---|---|---|---|
| *Candidate Interviews* | 0.050*** | 0.050*** | 0.050*** | 0.050*** |
| | (0.007) | (0.007) | (0.007) | (0.007) |
| *Unit Labour Cost* | 0.016 | 0.016 | 0.015 | 0.016 |
| | (0.016) | (0.017) | (0.016) | (0.016) |
| *Scale Intensive* | -0.106*** | -0.107*** | -0.107*** | -0.108*** |
| | (0.031) | (0.031) | (0.031) | (0.031) |
| *Science Based* | 0.025 | 0.026 | 0.025 | 0.025 |
| | (0.045) | (0.045) | (0.045) | (0.045) |
| *Supplier Dominated* | -0.129*** | -0.129*** | -0.129*** | -0.128*** |
| | (0.026) | (0.026) | (0.026) | (0.026) |
| *Size* | 0.000** | 0.000** | 0.000** | 0.000** |
| | (0.000) | (0.000) | (0.000) | (0.000) |
| *Exporter* | 0.084*** | 0.085*** | 0.084*** | 0.084*** |
| | (0.022) | (0.022) | (0.022) | (0.022) |
| *Industrial Group* | -0.016 | -0.015 | -0.016 | -0.018 |
| | (0.030) | (0.030) | (0.030) | (0.030) |
| *R&D* | 0.146*** | 0.146*** | 0.149*** | 0.150*** |
| | (0.023) | (0.023) | (0.023) | (0.022) |
| *Green R&D* | 0.080** | 0.081** | 0.082** | 0.075* |
| | (0.030) | (0.030) | (0.030) | (0.030) |
| *Editda* | -0.011 | -0.014 | -0.010 | -0.010 |
| | (0.010) | (0.010) | (0.010) | (0.010) |
| *High School by Employees* | 0.001 | 0.000 | 0.001 | 0.001 |
| | (0.021) | (0.021) | (0.021) | (0.021) |
| *Trade Unions* | 0.034 | 0.033 | 0.034 | 0.035 |
| | (0.024) | (0.024) | (0.024) | (0.024) |
| *% Graduates* | 0.008 | 0.008 | 0.008 | 0.008 |
| | (0.005) | (0.005) | (0.005) | (0.005) |
| *Age* | -0.000 | -0.000 | -0.000 | -0.000 |
| | (0.000) | (0.000) | (0.000) | (0.000) |
| *North* | 0.074** | 0.075** | 0.075** | 0.075** |
| | (0.028) | (0.028) | (0.028) | (0.028) |
| *South* | -0.033 | -0.030 | -0.032 | -0.032 |
| | (0.043) | (0.043) | (0.043) | (0.043) |
| Constant | 0.137 | 0.133 | 0.143 | 0.140 |
| | (0.186) | (0.187) | (0.186) | (0.186) |
| N. Obs. | 2636 | 2636 | 2636 | 2636 |

*Notes:* Reported coefficients are the results of the first stage Probit estimation for the endogenous variable STEM. Robust standard errors in parentheses. *** significant at the 1 % level; ** significant at the 5 % level; * significant at the 10 % level.

**Table A6** – Control Function - Second Stage

| | *InnoTot* | *CeI* | *ProcInno* | *ProdInno* |
|---|---|---|---|---|
| *Stem* | 0.576*** | 0.548*** | 0.427*** | 0.266*** |
| | (0.113) | (0.127) | (0.117) | (0.102) |
| *Trained Employees* | 0.00614*** | 0.00865*** | 0.00439*** | 0.00365*** |
| | (0.00152) | (0.00173) | (0.00149) | (0.00131) |
| *R&D* | 0.233*** | 0.0517 | 0.255*** | 0.400*** |
| | (0.0309) | (0.0366) | (0.0324) | (0.0293) |
| *Green R&D* | -0.0352 | 0.183*** | 0.000446 | -0.00483 |
| | (0.0351) | (0.0414) | (0.0368) | (0.0318) |
| *% Graduates* | -0.00425 | -0.0104** | -0.00221 | 0.00633* |
| | (0.00368) | (0.00496) | (0.00480) | (0.00380) |
| *High School by Employees* | -0.0342 | -0.0435* | -0.0237 | -0.00113 |
| | (0.0224) | (0.0259) | (0.0221) | (0.0191) |
| *Unit Labour Cost* | -0.0829*** | -0.0971*** | -0.0861*** | -0.0280** |
| | (0.0168) | (0.0195) | (0.0179) | (0.0143) |
| *Exporter* | 0.0461* | 0.0320 | 0.0306 | 0.136*** |
| | (0.0247) | (0.0279) | (0.0242) | (0.0217) |
| *Industrial Group* | -0.0322 | -0.0471 | 0.00438 | 0.00626 |
| | (0.0294) | (0.0347) | (0.0293) | (0.0257) |
| *Size* | -0.000563*** | -0.000752*** | -0.000538*** | -0.000160* |
| | (0.000123) | (0.000158) | (0.000167) | (0.0000958) |
| *Ebitda* | 0.0142 | 0.00165 | 0.0226* | -0.0119 |
| | (0.0125) | (0.0143) | (0.0134) | (0.00804) |
| *Scale Intensive* | 0.129*** | 0.178*** | 0.129*** | 0.00564 |
| | (0.0318) | (0.0379) | (0.0318) | (0.0281) |
| *Science Based* | -0.00887 | 0.0378 | -0.00498 | -0.000864 |
| | (0.0372) | (0.0520) | (0.0429) | (0.0390) |
| *Supplier Dominated* | 0.119*** | 0.146*** | 0.0907*** | -0.0262 |
| | (0.0300) | (0.0353) | (0.0296) | (0.0258) |
| *Trade Unions* | -0.00789 | 0.0457 | -0.0258 | -0.0475** |
| | (0.0259) | (0.0307) | (0.0254) | (0.0230) |
| *Age* | 0.000578 | 0.00123** | -0.000249 | 0.000415 |
| | (0.000504) | (0.000565) | (0.000465) | (0.000401) |
| *North* | -0.0201 | -0.0415 | -0.0399 | -0.00829 |
| | (0.0300) | (0.0342) | (0.0283) | (0.0243) |
| *South* | 0.0711 | 0.0816 | -0.0133 | 0.0801** |
| | (0.0444) | (0.0518) | (0.0435) | (0.0383) |
| *Constant* | 0.383** | 0.173 | 0.437** | 0.152 |
| | (0.195) | (0.221) | (0.199) | (0.166) |
| *cf (Stem)* | -0.271*** | -0.285*** | -0.188*** | -0.114* |
| | (0.0695) | (0.0780) | (0.0724) | (0.0624) |
| *cf (Trained Employees)* | -0.00621*** | -0.00856*** | -0.00410*** | -0.00352*** |
| | (0.00153) | (0.00174) | (0.00150) | (0.00132) |
| N. Obs. | 2636 | 2636 | 2636 | 2636 |
| Chi-Squared | 646.139 | 418.886 | 528.247 | 1421.652 |

*Notes:* Reported coefficients are second stage estimates of the control function approach. Robust standard errors in parentheses. The model includes the control functions (residuals) from the first stage estimations to account for endogeneity. Robust standard errors in parentheses. *** significant at the 1 % level; ** significant at the 5 % level; * significant at the 10 % level.

**Table A7** – Control Function - First stage Probit model for the endogenous variable STEM

|  | *STEM* |
|---|---|
| *R&D* | 0.441*** |
|  | (0.0585) |
| *Green R&D* | 0.281*** |
|  | (0.0923) |
| *% Graduates* | 0.0340** |
|  | (0.0151) |
| *High School by Employees* | 0.00443 |
|  | (0.0537) |
| *Unit Labour Cost* | 0.0756 |
|  | (0.0468) |
| *Exporter* | 0.212*** |
|  | (0.0554) |
| *Industrial Group* | -0.0563 |
|  | (0.0782) |
| *Size* | 0.00162** |
|  | (0.000661) |
| *Ebitda* | -0.0298 |
|  | (0.0321) |
| *Scale Intensive* | -0.312*** |
|  | (0.0788) |
| *Science Based* | 0.0220 |
|  | (0.125) |
| *Supplier Dominated* | -0.375*** |
|  | (0.0662) |
| *Trade Unions* | 0.0897 |
|  | (0.0655) |
| *Age* | -0.000968 |
|  | (0.00126) |
| *North* | 0.213*** |
|  | (0.0717) |
| *South* | -0.0719 |
|  | (0.114) |
| *Candidate Interviews* | 0.156*** |
|  | (0.0191) |
| Constant | -1.270** |
|  | (0.502) |
| N. Obs. | 2636 |
| Chi- squared | 337.131 |

*Notes:* Reported coefficients are the results of the first stage Probit estimation for the endogenous variable STEM. Robust standard errors in parentheses. *** significant at the 1 % level; ** significant at the 5 % level; * significant at the 10 % level.

**Table A8** – Control Function - First Stage Probit model for the endogenous variable Trained Employees

| | *Trained Employees* |
|---|---|
| *R&D* | 5.269*** |
| | (-1.610) |
| *Green R&D* | 6.886*** |
| | (-2.486) |
| *% Graduates* | 0.642** |
| | (0.293) |
| *High School by Employees* | 5.774*** |
| | (-1.367) |
| *Unit Labour Cost* | 2.564** |
| | (-1.180) |
| *Exporter* | -2.346 |
| | (-1.541) |
| *Industrial Group* | 3.330* |
| | (-1.996) |
| *Size* | 0.0295** |
| | (0.0121) |
| *Ebitda* | 1.712** |
| | (0.737) |
| *Scale Intensive* | -0.659 |
| | (-2.088) |
| *Science Based* | -4.324 |
| | (-3.170) |
| *Supplier Dominated* | -2.548 |
| | (-1.799) |
| *Trade Unions* | -1.015 |
| | (-1.687) |
| *Age* | 0.0109 |
| | (0.0342) |
| *North* | 4.704*** |
| | (-1.823) |
| *South* | 3.739 |
| | (-2.998) |
| *Training Reimbursement* | 4.024*** |
| | (0.437) |
| Constant | 31.91** |
| | (12.84) |
| N. Obs. | 2636 |
| F-stat | 12.172 |

*Notes:* Reported coefficients are the results of the first stage OLS estimation for the endogenous variable Trained Employees. Robust standard errors in parentheses. *** significant at the 1 % level; ** significant at the 5 % level; * significant at the 10 % level.

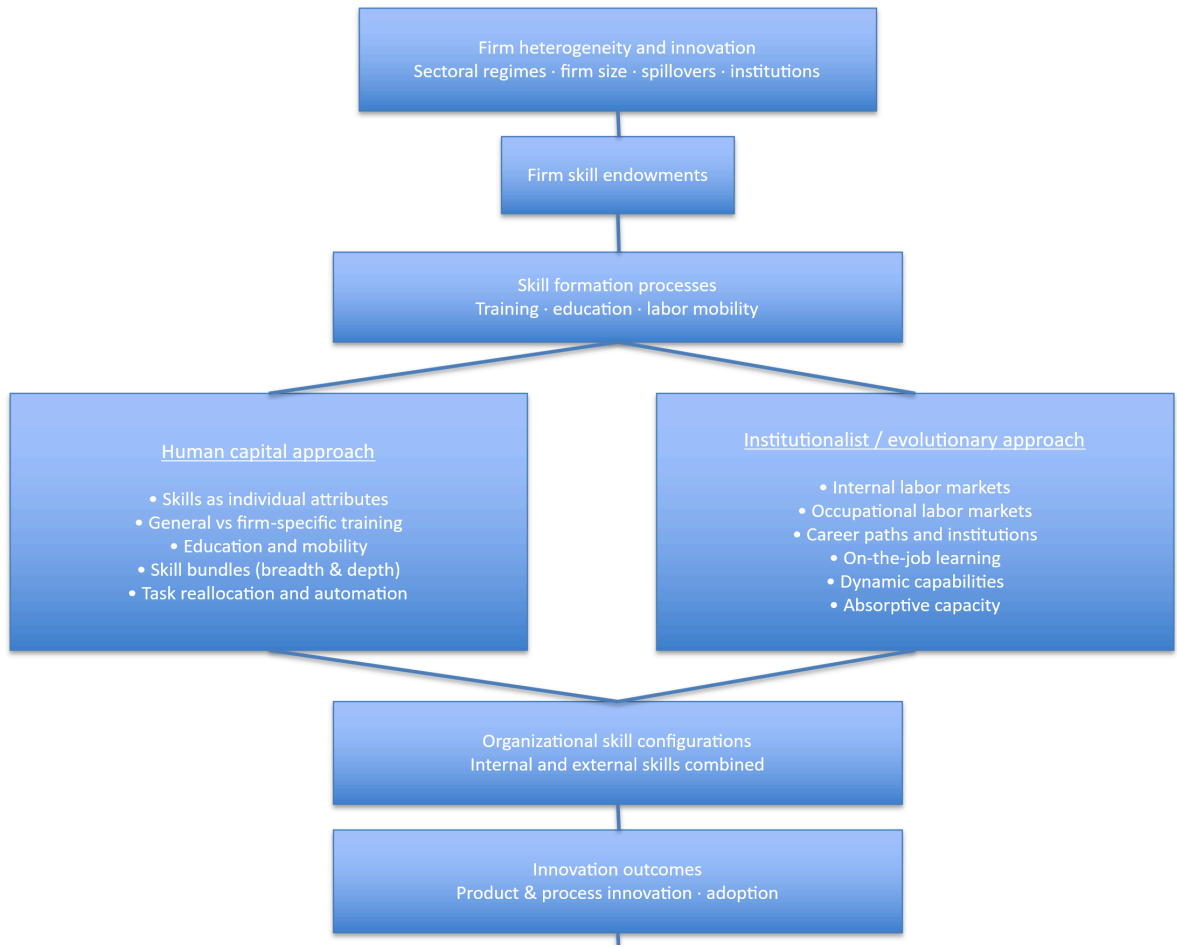